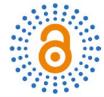

# A Possible Mechanism of DNA to DNA Transcription in Eukaryotic Cells: Endonuclease Dependent Transcript Cutout

## Gao-De Li

Chinese Acupuncture Clinic, Liverpool, UK
Email: gaode_li@yahoo.co.uk

## Abstract

**We previously proposed the existence of DNA to DNA transcription in eukaryotic cells, but the mechanism by which single-stranded DNA (ssDNA) transcript is produced and released from the genome remains unknown. We once speculated that the mechanism of DNA to DNA transcription might be similar to that of DNA to RNA transcription, but now we propose that endonuclease dependent transcript cutout may be a possible mechanism of DNA to DNA transcription, in which a copy of ssDNA fragment (transcript) between two nicks produced by nicking endonuclease is released from double-stranded DNA (dsDNA) region in the genome by an unknown ssDNA fragment releasing enzyme. The gap in the dsDNA will be filled through DNA repair mechanism. Occasionally, multiple copies of ssDNA transcripts could be produced through multiple rounds of cutout-repair-cutout cycle.**

## Keywords

**Plasmodium falciparum, Endonuclease Dependent Transcript Cutout, Mechanism, Eukaryotic Cell, Chromatin Conformation, Dynamic Genome Architecture, DNA to DNA Transcription, Nicking Endonuclease, ssDNA Transcript, Cell Cycle Progression, Chloroquine**

## 1. Introduction

Our recently published research findings obtained from *Plasmodium falciparum* strongly suggest the existence of DNA to DNA transcription in eukaryotic cells, which takes place before cell cycle progression into S phase. *i.e.* in G1 phase of the cell cycle [1] [2]. However, the mechanism by which a single-stranded DNA (ssDNA) transcript is produced and released from the genome remains unknown. We previously postulate that the mecha-nism of DNA to DNA transcription might be similar to that of DNA to RNA transcription, but except DNA repair-related limited DNA synthesis, no any other DNA synthesis in G1 phase of the cell cycle has been reported so far. Therefore, in this paper, we would like to present another possible mechanism for DNA to DNA transcription, which is different from that of DNA to RNA transcription and involves cutting DNA by endonuclease and DNA repair. Theoretically, both endonuclease activity and DNA repair function exist in G1 phase of the cell cycle, which could serve as the foundation for the mechanism.

## 2. Endonuclease Dependent Transcript Cutout Is Proposed as a Possible Mechanism of DNA to DNA Transcription

Endonuclease dependent transcript cutout is a novel idea proposed in this paper to understand the mechanism of DNA to DNA transcription, which is mainly based on conventional knowledge of endonuclease and DNA repair. It is well known that various restriction sites are distributed throughout the entire genome, surely including many restriction sites for nicking endonuclease that only cut one strand of a double-stranded DNA (dsDNA) to produce nicks in the dsDNA. The ssDNA fragment between the two nicks could be released from the dsDNA probably by an unknown enzyme that could be named as ssDNA fragment releasing enzyme. After the ssDNA fragment is released, a gap that is a temporary ssDNA region in dsDNA is formed, and will soon be filled through DNA repair mechanism using DNA polymerase and ligase. The released ssDNA fragment, like a RNA transcript, is a copy of a particular segment of DNA in the genome, and thus could be named as ssDNA transcript produced by means of DNA to DNA transcription.

Under certain conditions, the filled or repaired gap (DNA to DNA transcription region) in the dsDNA region of the genome might be repeatedly cut by the nicking endonuclease at the same restriction sites, forming a cut-out-repair-cutout transcription cycle. In each round of cycle one copy of ssDNA transcript is produced and released from the genome. To produce multiple copies of ssDNA transcripts, multiple rounds of such transcription cycles are needed.

No doubt, endonuclease dependent transcript cutout related DNA to DNA transcription should take place in G1 phase (possibly, in G2 phase as well) of the cell cycle, and before cell cycle progression into S phase, all the nicks and gaps in the dsDNA of the genome must be precisely repaired, and all the released ssDNA transcripts must be completely degraded so that the integrity of the genome is not affected during DNA replication. If DNA damage repair is not completed before cell cycle progression into S phase, the cell cycle arrest could occur.

Theoretically, if the restriction sites for nicking endonuclease are evenly distributed throughout the entire genome, the DNA to DNA transcription could take place in both coding and noncoding DNA regions of the genome. But we think that the DNA to DNA transcription might exclusively occur in noncoding DNA region of the genome because of three reasons, first, the vast majority of the mammalian genome does not code for proteins [3], and thus the chances that DNA to DNA transcription occurs in noncoding DNA region of the genome will be very high; second, the DNA to DNA transcription involves cutting DNA by endonuclease and DNA repair, which could go wrong to cause mutations if it occurs within coding DNA region; third, many genes have long or short introns which could be the places for nicks or gaps to occur and thus could protect the coding DNA regions from mutations. Conceivably, the endonuclease dependent transcript cutout is not only decided by the restriction sites, it is also restricted by DNA modifications, such as DNA methylation etc., and probably mainly decided by the genome architecture or chromatin conformation as some torsion or knots formed by the changes of three-dimensional structure of the genome at certain points during cell cycle progression require endonuclease dependent transcript cutout to sort them out.

Besides, the endonuclease dependent transcript cutout is a process similar to autotomy (self-injury) [4]: cutting off a piece of self DNA at a critical time of the cell cycle and the lost DNA part will be regenerated later. Therefore, in normal cell cycle only low levels of this transcription are maintained. Overdoing such transcription might cause cell cycle arrest or even cell death as a study showed that introduction of ssDNA fragments with diverse sequences into mammalian cells induced DNA damage as well as apoptosis signals [5]. The primary mechanism of chloroquine's antimalarial, anticancer, and immunomodulatory actions might be closely associated with excessive DNA to DNA transcription.

Taken together, endonuclease dependent transcript cutout may be a possible mechanism of DNA to DNA transcription in eukaryotic cells. Why is this self-injury transcription mechanism required for the cell cycle pro-gression? There are probably two main reasons for this. First, temporary ssDNA regions (gaps) in the genome are required for overcoming obstacles encountered by dynamic genome architecture or chromatin conformation during cell cycle progression. Second, the released ssDNA transcripts might have some important functions during cell cycle progression, such as being involved in the regulation of gene expression (most likely, down-regulating cell cycle control gene expression), forming certain complexes with other molecules, and generating intrinsic primers for natural site-directed mutagenesis [6].

## 3. Evidence to Support the Endonuclease Dependent Transcript Cutout as a Possible Mechanism of DNA to DNA Transcription

To explore the mechanisms of DNA to DNA transcription, we have reviewed some publications about ssDNA in isolated nuclear DNA so that a clue of ssDNA transcript released through DNA to DNA transcription could be found [7]-[11]. Generally, ssDNA amounts to 1% - 2% of the total nuclear DNA samples isolated from various eukaryotic cells [7] [10]. The origin of ssDNA might be helpful in identification of ssDNA transcripts from the isolated ssDNA. To date, all the articles we reviewed clearly showed that majority of ssDNA was derived from DNA to RNA transcription regions of the genome through selective endogenous nuclease attacks presumably at an early stage of the DNA purification procedure [8] [9]. In another word, these articles indicate that the majority of ssDNA contained in the isolated nuclear DNA from various eukaryotic cells does not really exist in living cells, and was artificially produced during isolation procedure.

As for the origin of small fraction of ssDNA that was not derived from DNA to RNA transcription regions, the majority of the articles we reviewed didn't mention it at all. Luckily, we eventually found one article, in which the origin of a small faction of ssDNA was clearly mentioned: "consists of self-reassociating, moderately repeated DNA sequences, mainly derived from noncoding regions of the cell genome" [10]. We think that this small fraction of ssDNA with repeated sequences might be produced by endonuclease dependent transcript cutout, and is just the ssDNA transcript we are looking for.

Another evidence to support our DNA to DNA transcription mechanism comes from two articles, which showed that the amount of single strandedness of DNA in the human diploid fibroblasts and HeLa cells started to increase from G1, reached a maximum during S phase, and decreased as the cells enter G2 [12] [13]. The maximum of single strandedness in S phase is understandable because it has been reported to be derived from the forks of replicating DNA [14] [15]; as for the origin of the single strandedness of DNA detected in G1 phase, we think that at least some ssDNA fragments or molecules might be produced by endonuclease dependent transcript cutout.

Discovery of ssDNA regions within dsDNA helix in G1 phase is also an indirect evidence to support our idea. One article showed that DNA isolated from Chinese hamster ovary cells synchronized in G1 phase contained ssDNA regions separated by a distance of 100 μm. The author emphasized that the ssDNA regions in the G1 cells were not to be the result of a low level of DNA replication nor to be an artifact of the isolation procedure [16]. The mechanism by which these ssDNA regions were produced was not mentioned in the paper, but it is possible that they might be produced by endonuclease dependent transcript cutout and had not yet been completely repaired when they were isolated from the nucleus.

To date, except certain DNA repair that involves a limited DNA synthesis, no any other DNA synthesis in G1 phase has been reported. Thus, the only theoretical foundation for endonuclease dependent transcript cutout, a possible mechanism for DNA to DNA transcription proposed by us in this paper, will be the fact that both endonuclease activity and DNA repair function exist in G1 phase of eukaryotic cell cycle.

## 4. Implications of the Mechanism of DNA to DNA Transcription

The significance of endonuclease dependent transcript cutout proposed as a possible mechanism for DNA to DNA transcription in this paper is obvious as the mechanism belongs to one of fundamental functions of the genome. DNA to RNA transcription mainly deals with the gene expression, while DNA to DNA transcription is associated with the dynamic genome architecture during cell cycle progression, and also involved in the regulation of gene expression and generating intrinsic primers for natural site-directed mutagenesis in eukaryotic cells etc.. Therefore, once the mechanism is proven to be true, it will certainly enrich our knowledge about the functions of the genome, which will accelerate the progress in life sciences.

## 5. Conclusion

In this paper, we proposed that endonuclease dependent transcript cutout may be a possible mechanism of DNA to DNA transcription in eukaryotic cells, which is based on conventional knowledge of endonuclease and DNA repair. To date, some indirect evidence from publications has been found to support the idea, but they are not enough. Therefore, design of appropriate experiments to validate the mechanism is both necessary and worthwhile.

## Conflict of Interest

The author declares that there is no conflict of interest regarding the publication of this paper.

## References

[1]  Li, G.D. (2016) Certain Amplified Genomic-DNA Fragments (AGFs) May Be Involved in Cell Cycle Progression and Chloroquine Is Found to Induce the Production of Cell-Cycle-Associated AGFs (CAGFs) in *Plasmodium falciparum*. *Open Access Library Journal*, **3**, e2447. http://dx.doi.org/10.4236/oalib.1102447

[2]  Li, G.D. (2016) DNA to DNA Transcription Might Exist in Eukaryotic Cells. *Open Access Library Journal*, **3**, e2665. http://dx.doi.org/10.4236/oalib.1102665

[3]  Smith, N.G., Brandström, M. and Ellegren, H. (2004) Evidence for Turnover of Functional Noncoding DNA in Mammalian Genome Evolution. *Genomics*, **84**, 806-813. http://dx.doi.org/10.1016/j.ygeno.2004.07.012

[4]  Clause, A.R. and Capaldi, E.A. (2006) Caudal Autotomy and Regeneration in Lizards. *Journal of Experimental Zoology Part A: Comparative Experimental Biology*, **305**, 965-973.

[5]  Nur-E-Kamal, A., Li, T.K., Zhang, A., Qi, H., Hars, E.S. and Liu, L.F. (2003) Single-Stranded DNA Induces Ataxia Telangiectasia Mutant (ATM)/p53 Dependent DNA Damage and Apoptotic Signals. *The Journal of Biological Chemistry*, **278**, 12475-12481. http://dx.doi.org/10.1074/jbc.M212915200

[6]  Li, G.D. (2016) "Natural Site-Directed Mutagenesis" Might Exist in Eukaryotic Cells. *Open Access Library Journal*, **3**, e2595. http://dx.doi.org/10.4236/oalib.1102595

[7]  Hanania, N., Schaool, D., Poncy, C., Tapiero, H. and Harel, J. (1977) Isolation of Single Stranded Transcription Sites from Human Nuclear DNA. *Cell Biology International Reports*, **1**, 309-315. http://dx.doi.org/10.1016/0309-1651(77)90060-1

[8]  Hanania, N., Shaool, D. and Harel, J. (1982) Isolation of a Mouse DNA Fraction Which Encodes More Informational Than Non Informational RNA Sequences. *Molecular Biology Reports*, **8**, 91-96. http://dx.doi.org/10.1007/BF00778510

[9]  Leibovitch, S.A., Tapiero, H. and Harel, J. (1977) Single-Stranded DNA from Oncornavirus-Infected Cells Enriched in Virus-Specific DNA Sequences. *Proceedings of the National Academy of Sciences of the USA*, **74**, 3720-3724. http://dx.doi.org/10.1073/pnas.74.9.3720

[10] Leibovitch, S.A. and Harel, J. (1978) Active DNA Transcription Sites Released from the Genome of Normal Embryonic Chicken Cells. *Nucleic Acids Research*, **5**, 777-787. http://dx.doi.org/10.1093/nar/5.3.777

[11] Leibovitch, S.A., Tichonicky, L., Kruh, J. and Harel, J. (1981) Single-Strandedness of the Majority of DNA Sequences Complementary to mRNA-Coding Sites Isolated from Rat Hepatoma Tissue Cultured Cells. *Experimental Cell Research*, **133**, 181-189. http://dx.doi.org/10.1016/0014-4827(81)90368-2

[12] Collins, J.M. (1977) Deoxyribonucleic Acid Structure in Human Diploid Fibroblasts Stimulated to Proliferate. *Journal of Biological Chemistry*, **252**, 141-147.

[13] Collins, J.M., Berry, D.E. and Cobbs, C.S. (1977) Structure of Parental Deoxyribonucleic Acid of Synchronized HeLa Cells. *Biochemistry*, **16**, 5438-5444. http://dx.doi.org/10.1021/bi00644a006

[14] Wanka, F., Brouns, R.M., Aelen, J.M., Eygensteyn, A. and Eygensteyn, J. (1977) The Origin of Nascent Single-Stranded DNA Extracted from Mammalian Cells. *Nucleic Acids Research*, **4**, 2083-2097. http://dx.doi.org/10.1093/nar/4.6.2083

[15] Bjursell, G., Gussander, E. and Lindahl, T. (1979) Long Regions of Single-Stranded DNA in Human Cells. *Nature*, **280**, 420-423. http://dx.doi.org/10.1038/280420a0

[16] Henson, P. (1978) The Presence of Single-Stranded Regions in Mammalian DNA. *Journal of Molecular Biology*, **119**, 487-506. http://dx.doi.org/10.1016/0022-2836(78)90198-5

**Open Access Library**

**Warmly welcome your paper submission to OALib Journal!**

- Publication on a daily basis
- 9 subject areas of science, technology and medicine
- Fair and rigorous peer-review system
- Fast publication process
- Article promotion in various social networking sites (LinkedIn, Facebook, Twitter, etc.)
- Widely-targeted and multidisciplinary audience to read your research

Submit Your Paper Online: **Click Here to Submit**
Contact Us: **service@oalib.com**